# Infrared narrow band gap nanocrystals: recent progresses relative to imaging and active detection

Charlie Greboval[1], Simon Ferre[2], Vincent Noguier[2], Audrey Chu[1,3], Junling Qu[1], Sang-Soo Chee[1], Gregory Vincent[3], Emmanuel Lhuillier[1*]

[1] *Sorbonne Université, CNRS, Institut des NanoSciences de Paris, INSP, F-75005 Paris, France*

[2] *New Imaging Technologies SA, 1 impasse de la Noisette 91370 Verrières le Buisson, France*

[3] *ONERA−The French Aerospace Lab, 6, Chemin de la Vauve aux Granges, BP 80100, F-91123 Palaiseau, France*

**ABSTRACT:**

Current technologies for infrared detection have been based on epitaxially grown semiconductors. Here we review some of the recent developments relative to colloidal nanocrystals and their use as building blocks for the design of low-cost infrared sensors. We focus on HgTe nanocrystals which appear as the only material leading to infrared photoconductivity and ultra-broad spectral tunability: from the visible to the long-wave infrared. We review some of the important results which demonstrated that colloidal nanocrystals can be compatible with air stable operations, fast detection, and strong absorption. We discuss the recent progresses relative to multipixel devices and show results obtained by coupling short-wave infrared nanoparticles with CMOS circuits to achieve video rate VGA format imaging. In particular we present that nanocrystals are a promising material for long range (>150 m) active detection in both continuous wave and pulsed mode with a time resolution down to 10 ns.

## INTRODUCTION

Current technologies for infrared (IR) sensing are driven by epitaxially grown semiconductors. The high maturity of technologies such as InGaAs, InSb, HgCdTe, or type II superlattices [1] has led to exceptional performances which include high signal-to-noise ratio, large format focal plane arrays (FPAs), and fast detection. However, the prospect of costs reduction is extremely limited. This makes that IR sensing is mostly limited to two niche markets, including the defense and astronomy.

Thermal detectors such as pyrometers and bolometers offer room temperature operation and reduced fabrication costs, but signal-to-noise ratio is lower and time response is slower compared with epitaxially grown detectors. Among these technologies, the bolometer is certainly the one achieving the best imaging performances however, it operates in the long wave infrared (LWIR) which let both short-wave IR (SWIR) and mid-wave IR (MWIR) without real low-cost alternative. Extremely cheap (a few hundred euros) focal plane arrays (FPAs) are available but the number of pixels and the frame rate are restricted, leading to low quality images.

There is an alternative technology if the latter can combine the efficiency of quantum sensors (speed and spectral tunability) with the low costs of thermal detectors. Such detector might be useful for applications in night driving, industrial vision for fast sorting, and LIDAR detection. All these applications require a cost break down by at least a factor of 5 compared to current IR sensors. When it comes to low cost, organic-electronic materials are often seen as a possible candidate, but they are ineffective beyond 1 µm.

Several materials have been proposed as alternative candidates such as graphene [2,3] or black phosphorus [4,5]. However, colloidal nanocrystals are particularly those for which most of the efforts have been continuously developed over the past 10 years: switching from the proof of concept for mid IR detection [6] to a high performance versatile platform for IR optoelectronics [7,8].Their interest has first been raised by their ability to emit light. This has led to their integration into displays as narrow green and red sources. Combining the broad range of materials that can be synthesized under colloidal form with quantum confinement it is possible to extend the optical spectrum of such nanocrystals from the UV to the THz range [9].

In the IR range, nanocrystals have drawn tremendous attention for the design of solar cells to collect the near IR part of the solar spectrum. In particular the observation of a low energy threshold for multiexciton generation in nanocrystals offered a path to overcome Shockley Queisser limitation, compared to the bulk [10,11]. This is only later (≈2005) that the use of nanocrystals for IR sensing have begun to attract interest [12,13]. Nanocrystals combine reasonably low fabrication costs and benefit from the processability of organic electronics. They can be spin-coated or ink-jet printed, which is of utmost interest to address the complex and expensive hybridization step to the read-out circuit.

In this paper we review some of the recent developments relative to the integration of colloidal nanocrystals for IR sensing [14–18]. We chose to focus on mercury chalcogenides [8,19,20] which currently appear as the only material allowing all IR ranges from SWIR to the THz. We try to cover every aspect from the material growth, to concepts introduced to push performance at the single pixel level and finally recent demonstrations relative to image sensing.

**DISCUSSION**

The first colloidal nanocrystals with interband absorption that were developed where based on lead chalcogenides. They are well suited to be used as solar cell absorbers thanks to their tunable band gap around 1.2 eV [21,22]. However, because of their bulk band gap, lead chalcogenides are not appropriate to explore wavelengths above 4 µm [23] at least under colloidal form [24]. Narrower band gap materials or semimetals appear to be better suited to explore longer wavelengths. By far, the colloidal synthesis of nanocrystals has been focused on II-VI semiconductors and especially CdSe [25]. Consequently, semimetals consisting of II-VI materials where looking the most promising to reach the IR taking advantage of the chemical knowledge previously developed. This was pledging for the development of mercury chalcogenide compounds [13,26,27], which was further reinforced by their past extensive use in the IR detection field.

**Mercury Chalcogenides nanocrystals, the most tunable platform for infrared optoelectronics**

While CdSe nanocrystals have been widely investigated for several years, only a few work have been done on HgTe before 2010 [13]. In 2010, Keuleyan et al.[28] have proposed the first colloidal synthesis facilitating mid-IR absorption in nanocrystals, see Figure 1b. This first material was strongly aggregated as shown transmission electronic microscopy image, see Figure 1a. This aggregation explains the poorly defined and smooth excitonic feature of the absorption spectrum. Later, the synthesis was improved to reduce aggregation of the nanocrystals, see Figure 1c. As objects were better defined, the band edge absorption feature was also sharper, see Figure 1d. Operating the same growth procedure in more dilute conditions decreased polydispersity, see Figure 1e. We obtained, as a result, an "atom-like" spectrum and up to six excitonic transitions can be observed in the absorption spectrum, see Figure 1f.

The synthesis of 2D HgTe colloidal quantum wells [28,29] demonstrated an improved control of the material growth see Figure 1g. The obtained material is often called nanoplatelets (NPLs) [30,31] and offers interesting features: the growth mechanism makes that the only confined direction presents no roughness and thus atomic control is achieved. This gives rise to the narrowest optical feature achieved for nanocrystals due to the lack of inhomogeneous broadening. Currently, the thinnest material is only 3 monolayers thick, which leads to an optical feature in the near IR (800-1000 nm), that can expanded to the short-wave IR by the growth of a shell [32,33]. Because of the semimetal nature of the bulk material, all the energy of the optical transition is due to quantum confinement: this makes HgTe NPLs one of the most confined nanocrystal material with approximately 1.5 eV of confinement.

It is also possible to explore extremely reduced confinement in this material. The Bohr radius of HgTe [34] is around 40 nm, which requires to grow large nanoparticles. This was presented by Goubet et al who have proposed a synthetic method where the size of the nanoparticles can be tuned from 5 nm and up to 1 µm [9], see Figure 1i. The obtained nanoparticles can absorb from long wave IR up to the THz range [9,35] (60 µm for the absorption peak and up to 200 µm for the absorption edge, see Figure 1j).

Last, it is worth also mentioning that heterostructures (core-shell objects) based on mercury chalcogenides have also been demonstrated [36–38], even though several challenges are remaining to further enhance the luminescence efficiency and decrease the non-radiative decay path of the exciton [18].

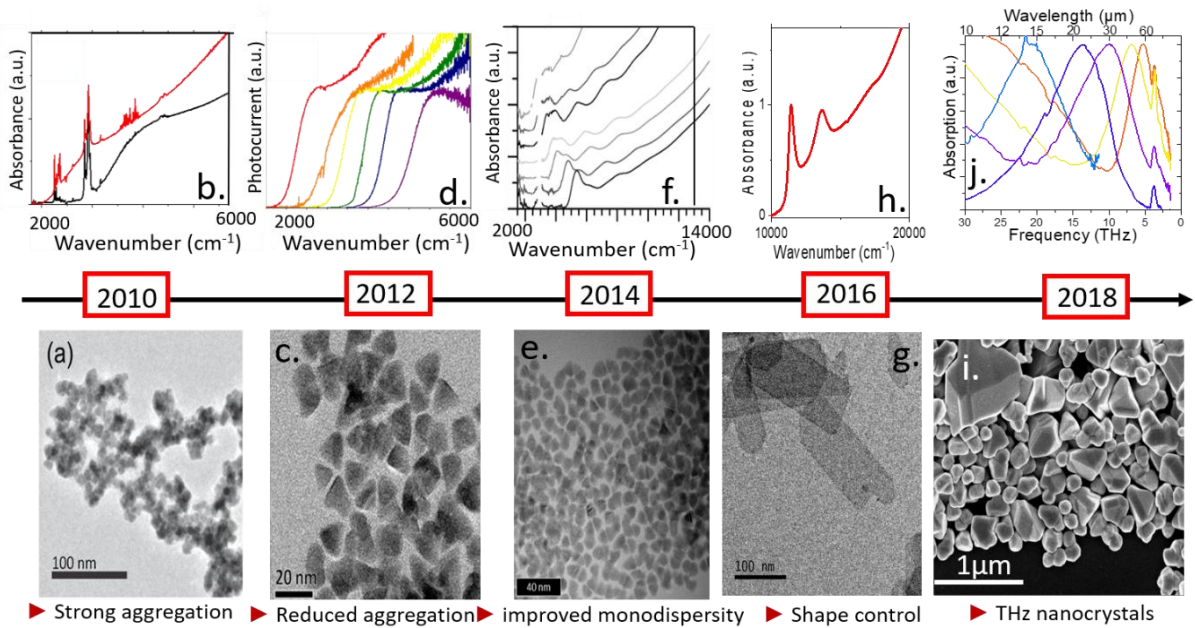

*Figure 1 TEM (a) and associated absorption spectra (b) of the fisrt nanocrystals absorbing in the mid-IR [6]. TEM (c) and absorption spectra (d) of HgTe mid-IR nanocrystals with reduced agregation. Adapted from ref [28] . Copyright (2011) American Chemical Society. TEM (e) and absorption spectra (f) of HgTe mid-IR nanocrystals with improved monodispersity. Adapted with permission from ref [29] . Copyright (2014) American Chemical Society. TEM (h) and associated absorption spectra (g) of HgTe colloidal 2D quantum wells. Adapted with permission from ref [30]. Copyright (2016) American Chemical Society. TEM (i) and associated absorption spectra (j) of THz HgTe nanocrystals. Adapted with permission from ref [9] . Copyright (2016) American Chemical Society.*

The principle of IR absorption in colloidal nanocrystals can be classified by two different mechanisms: interband transition in a narrow band gap/semi metal material or intraband absorption. The latter requires degenerate doping, which has been a synthetic challenge for long. Significant progresses have been realized over the recent years, to obtain doped nanoparticles via introduction of extrinsic impurities within the NCs [39–41] or within the nanocrystal array [42], non-stoichiometry of the material ($Cu_{2-x}S$ [43], $(Bi;Sb)_2Te_3$, HgSe [35,44]), metal functionalization [45,46], and functionalization by redox molecules [47,48] or surface dipole functionalization [49–53].

In the case of mercury chalcogenides, the doping magnitude will tune the observed IR transition. When the size of particles is small, confinement reduces the doping and this is mostly interband absorption that is observed, see Figure 2a and d. When confinement is reduced but is still present (typically in the 3-12 µm range), intraband absorption, which is the 0D analog of the intersubband absorption in quantum wells, is observed [44,51,54,55], see Figure 2b and d. When doping is vanishing or doping level is very high, the density of state becomes dense and the optical feature acquires more and more a metallic nature, leading to the observation of plasmonic transitions [9,53,56]. This unique combination with various types of transitions and their tunability thanks to doping and quantum confinement make that mercury chalcogenides are the most tunable nanocrystals. The energy of the first excitonic feature can be tune by almost two orders of magnitude. For sake of comparison, in CdSe, quantum confinement offers not more than 30% of tunability for the energy of the excitonic feature.

**The electronic structure of mercury chalcogenides nanocrystals**

A critical step for the integration of these nanocrystals into devices, especially for diode type devices, is the determination of their electronic spectrum in absolute energy scale. The band diagram of bulk HgTe is already unusual with an inverted band structure [57] (i.e. the band with a $\Gamma_6$ symmetry, which is usually the conduction one is below the Fermi level and is deep in the valence band). The other distinction already mentioned is the semi-metal nature of this material which is used

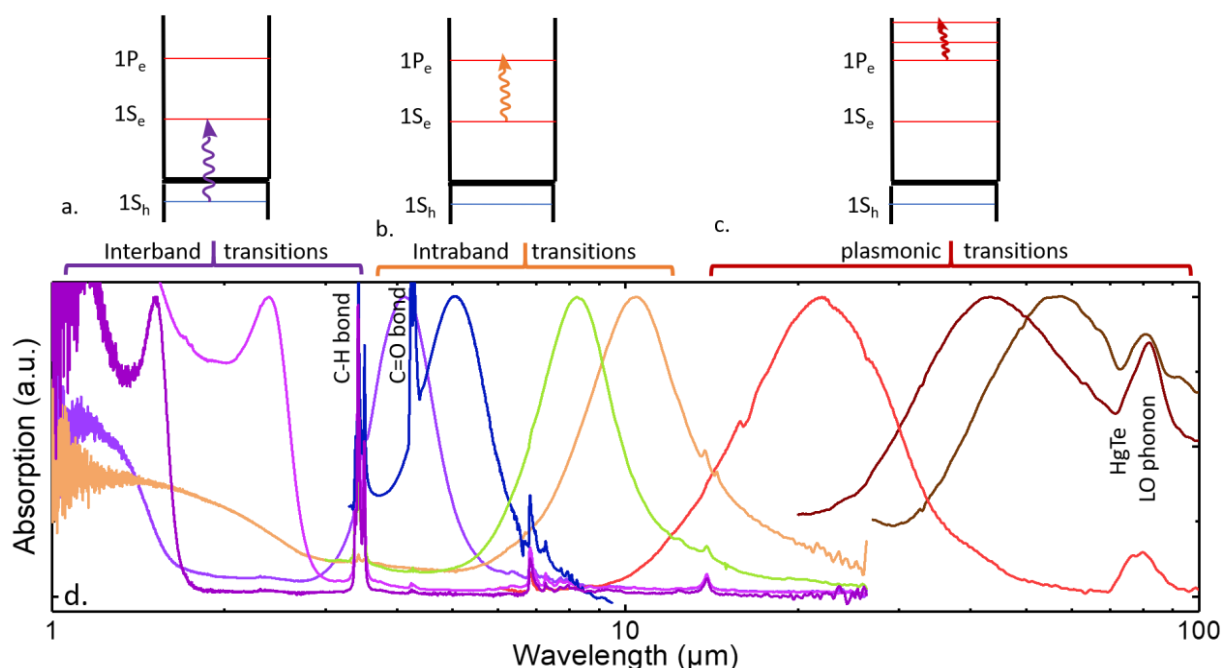

*Figure 2 (a), (b) and (c) are respectively the schemes for interband, intraband and plasmonic transitions in nanocrystals. d. Absorption spectra for mercury chalcogenide (HgSe and HgTe) nanocrystals with various sizes. The figure is adapted from ref [17].*

for THz absorption thanks to vanishing confinement. The optical absorption in this material occurs between the two bands with $\Gamma_8$ symmetry [58,59], see **Erreur ! Source du renvoi introuvable.**a. Note that quantum confinement in HgTe nanocrystals is used to tune the energy of the absorption cut-off similarly to the Cd content for bulk HgCdTe alloys.

In contrast to the bulk original band structure, HgTe nanocrystals add two other specificities which are the quantum confinement (**Erreur ! Source du renvoi introuvable.**b) and the dependence on the spectrum with surface chemistry (**Erreur ! Source du renvoi introuvable.**c). These different effects combined make this is currently impossible to predict *a priori* the exact energy level of each band for a given nanocrystal. This is why the exact energy has to be measured. Two main methods can be used: electrochemistry [29,53,60–62] or more conventionally for semiconductors: X-ray photoemission spectroscopy. The work function of the material has been found to be 4.6±0.1 eV and poorly depends on the size. It is interesting to notice that surface chemistry can induce p- and n-type nature to a given material and so to a given size. This was for example used in the design of p-n junction [63]. In the case of HgTe, the determination of the electronic spectrum was used to build by design a unipolar barriers used to filter the dark current [60] in a diode geometry.

**Detection-oriented performance**

Integration of nanocrystals into devices to replace historical technologies not only requires achieving

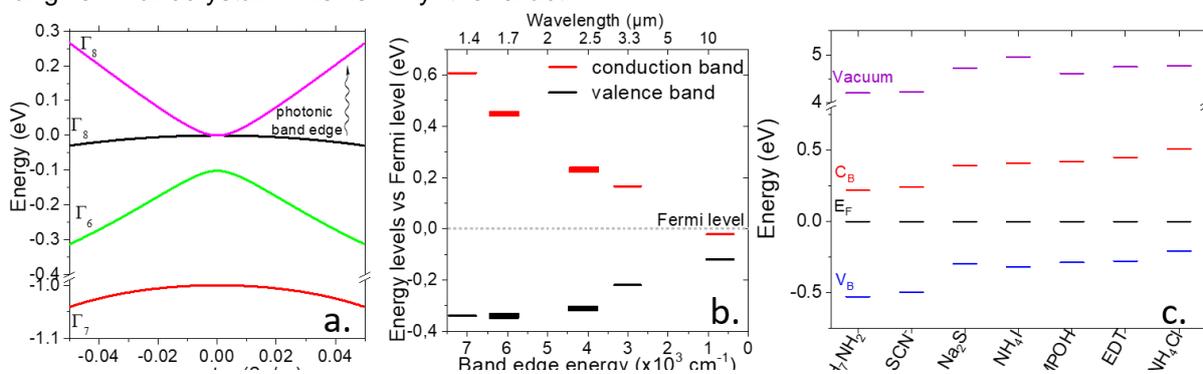

*Figure 3 (a) Band structure of bulk HgTe. Energy levels of the valance band, conduction band, and vacuum with respect to the Fermi energy set at 0 eV, (b) for various sizes of HgTe nanocrystals capped with ethanedithiol and (c) for HgTe with a band edge energy corresponding to 6000 cm$^{-1}$ and various capping surface chemistry. Part (b) and (c) are respectively adapted with permission from ref [61] and [62]. Copyright (2018 and 2019) American Chemical Society.*

high signal-to-noise ratio, but also imposes to provide similar temporal stabilities and fast time response. For long these issues have not been still addressed.

Recently we started to tackle this issue. HgTe nanocrystal arrays are intrinsically porous which makes the material more exposed to oxidation by the environment. We indeed observe that the conductance of a film stored in air rises by a factor >50 over a week, see Figure 4a. The material is also sensitive to temperature because of its low temperature growth (60 to 120 °C depending on the targeted size). Once exposed to high temperature the nanoparticles sinter, which increases the delocalization length, reduces the optical band gap and increases the dark current. Thus, we proposed a room temperature deposition of a multilayer system which is air (PVA) and water (PMMA and PVDF) repealing. An encapsulated film achieved

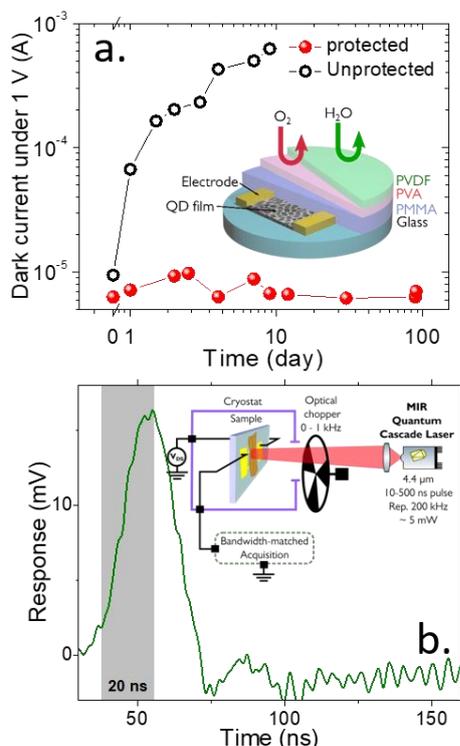

*Figure 4 (a) Dark current as a function of time spent in air for a HgTe CQD film with and without protective layers. Time zero corresponds to the sample in the air-free glove box. Inset is a scheme of the HgTe CQD film device with protective layers. The figure is adapted with permission from ref [65] . Copyright (2018) American Chemical Society. (b) Typical HgTe nanocrystal film response to a laser-limited 20 ns Quantum Cascade laser pulse at 4.4 µm. The inset is a scheme of the measurement setup with a mid-infrared Quantum Cascade laser. The figure is adapted with permission from ref [66]. Copyright (2018) American Chemical Society.*

stable dark current over three months. More work will have to be done in this direction.

Another challenge about the potential use of nanocrystals lies in transport and time response optimization. Transport in nanocrystal arrays occurs via a hopping mechanism, typically a nearest neighbor hopping at room temperature. First, this mechanism requires essentially a ligand exchange procedure. During this process, the native long ligands grafted onto the nanocrystal surface behaving as a tunnel barrier, are stripped and replaced by shorter organic molecules or even inorganic ions [64,65]. Such procedure has been extremely useful to improve the carrier mobility [66] (now around $1 cm^2.V^{-1}s^{-1}$ [67]). Such long ligands (and tunnel barriers) have presumably prevented fast detection. In addition, for a long time, the time response of mid-IR nanocrystals was probed in a non-appropriate way using high energy pulsed lasers. Fast detection times were observed but may also have been the result of hot electrons cooling. Using a quantum cascade laser resonant with the band edge of MWIR nanocrystals (see the inset of Figure 4b), we have recently demonstrated that the electrical time response of HgTe thin films can be no longer than 20 ns [68], see Figure 4b.

**Photon Management in HgTe**

The previously mentioned low carrier mobility leads to a short carrier diffusion length (below 100 nm), which is typically 1 order of magnitude smaller than the material absorption depth (a few µm) [59]. Consequently, thin nanocrystal films only absorb a limited part of the incident light. As a sake of example, a film of HgTe nanocrystals which is 200 nm thick and with an excitonic feature at 2.5 µm only absorb 12% of the incident light. Building thick films may increase absorption, but the charge collection will not be improved due to this short carrier diffusion length. In 2019, a significant effort has been done to develop a new film deposition method which combines better inter-nanocrystal coupling and allows the deposition of thicker films [66,67]. Films with thicknesses above 500 nm can now be routinely obtained and absorb around 40% of the incident light, see Figure 5a.

To further enhance the absorption of a nanocrystal film, it has been proposed to introduce plasmonic resonators [69–76]. Our approach to this issue is to modify the plane wave propagation of the incident light to generate a guide mode in the film. To do so a grating which spacing is roughly given by the wavelength divided by the medium dielectric constant is placed below the film. We proposed a device which combines a back-side reflector with

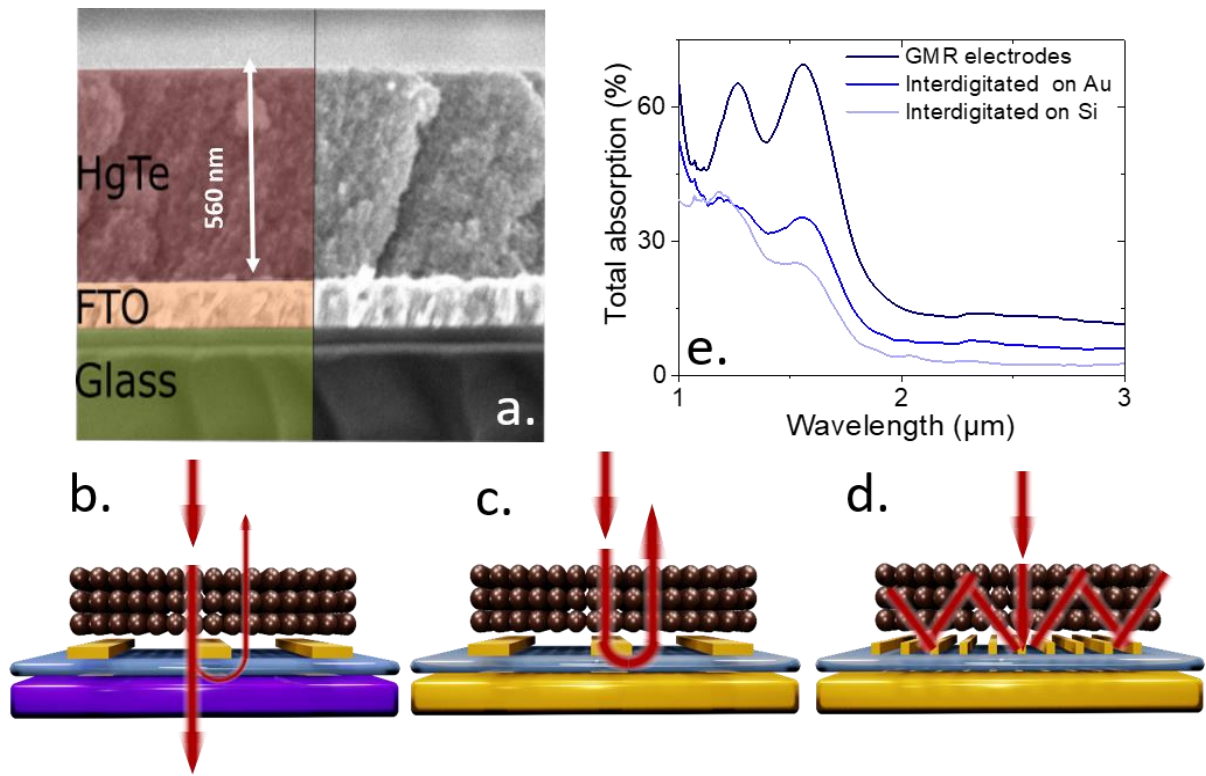

*Figure 5 (a) SEM cross-section of the diode made from a thick strongly absorbing film of HgTe nanocrystals, adapted from ref [69]. (b-d). Schematics of the cross section of (b) the conventional interdigitated gold electrodes deposited on a Si/SiO$_2$ substrate, (c) the interdigitated electrodes made of gold contact deposited on an Au/SiO$_2$ substrate and (d) the interdigitated gold electrodes on an Au/SiO$_2$ substrate designed to induce Guided mode resonance. Red arrows correspond to light paths. (e) Simulated absorption spectra for electrodes described in part (b-d). The figure is adapted with permission from ref [71] . Copyright (2019) American Chemical Society.*

such guided mode resonator [77]. The back-side mirror is here to generate two passes for the incident light while the grating generates multi passes of the light as explained on the scheme of Figure 5b-d. The obtained device can absorb 70% of the total incident light and even 100% of the light propagating along the transverse magnetic (TM) mode. In such a device a lot of attention have been paid to locate the absorption within the semiconductor and not in the metal to avoid thermal losses. Here around 80% of the absorption occurs in the semiconductor, this is good enough to enhance absorption but also offers interesting perspectives for further improvements.

**Short Wave Infrared imaging using HgTe nanocrystals**

All previously discussed results were focused on single pixel devices. It is worth pointing that the interest in nanocrystals is not only generated by the reduced fabrication cost but possibly also by their ease to be hybridized to the read-out circuit. Thanks to their solution processability, direct deposition of the nanocrystal film onto the read-out circuit should be possible. There have been several demonstrations of this concept either using PbS nanocrystals in the near IR [78] or using HgTe in the MWIR [79,80]. Here we rather focus on the SWIR as a possible alternative to InGaAs. Our first attempt was very basic and based on a home-made 10x10 pixel array, see Figure 6a and b. The deposition of a HgTe ink leads to photoconductive and strongly absorbing film as described in Figure 5a. This reduced-size focal plane array was then used as an IR laser beam profiler, see Figure 6c.

To achieve a higher image quality, the nanocrystal film was then deposited on a VGA format read-out integrated circuit (ROIC) by New Imaging Technologies. One can follow the recent progresses achieved while coupling such absorbing nanocrystal films to these ROICs in Figure 6c-d. First demonstrator (Figure 6c) was requiring direct illumination by a laser, suggesting a poor external quantum efficiency. It is now possible (Figure 6d) to make passive imaging of a scene in the near IR (i.e. visible part of the sun light is removed by a high pass filter).

**Use of nanocrystals for active detection**

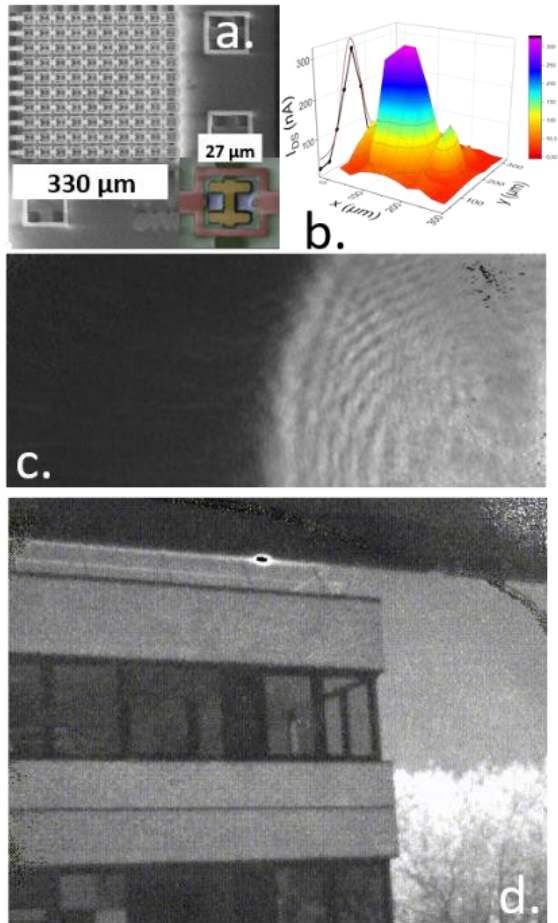

*Figure 6 (a) SEM image of a pixel matrix used as 10x10 pixel array of electrodes to design a first generation of HgTe nanocrystal based focal plane array. Inset is a SEM image of the matrix with fake colors corresponding to different layers: yellow is the internal gold contacts (columns), green is the insulating silica and orange is the external gold contacts (lines). (b) Current mapping of the photocurrent once the matrix is illuminated with a focused laser spot. The gaussian shape has a FWMH of 75 µm. (c) Image of a 1.55 µm laser spot for first attempt of HgTe nanocrystal film connected to a CMOS ROIC. The figure is adapted with permission from ref [62] . Copyright (2019) American Chemical Society. (d) Image from a HgTe nanocrystal film connected to CMOS ROIC, with no external lightning (i.e. passive imaging) and while the visible part of the spectrum has been removed using a high pass filter.*

To finish this discussion, we would like to mention another promising result for the use of nanocrystals in active detection. PbS nanocrystals based on solar cells have actually been designed to efficiently absorb the near IR part of the solar spectrum, but they are used as broad-band absorbers in solar cells. We have actually revisited this concept and operated them as sensors for near IR active detection [81]. Such a detector presents a high detectivity above $10^{12}$ Jones at room temperature and a fast time response (10 ns rise time – 1 µs decay time). We demonstrate that they can detect back scattered light between two building spaced by 85 m, see Figure 7a and b. We also show that they can be used for time-of-flight measurements, see Figure 7c. This is promising for their use in LIDAR detection as long as similar performance can be preserved at telecom wavelengths.

# CONCLUSION

We have discussed some of the recent developments of IR nanocrystals and their use for IR sensing including for imaging and active detection. 10 years ago, the challenge was to grow nanocrystals absorbing in the IR. Material growth is now mature, and the electronic structure of the material is mostly known. Great progresses have also been obtained for the doping of the nanoparticles which has open the way for the use of intraband absorption. Nanocrystals can achieve air stable (>3 months), fast (down to 20 ns) operation and absorb most (>70%) of the incident light. Large format imaging is also possible with limited dead pixels and edge effect. More sophisticated operation mode based on active detection and time gated operation appears to be also possible.

# ACKNOWLEDGMENT

EL acknowledges the financial support of the European Research Council (ERC) starting grant (blackQD – n°756225). We thank Agence Nationale de la Recherche for funding through grant IPER-nano2, Copin, Graskop and Frontal. This work was supported by French state funds managed by the ANR within the Investissements d'Avenir programme under reference ANR-11-IDEX-0004-02, and more specifically within the framework of the Cluster of Excellence MATISSE. We acknowledge the use of clean-room facilities from the "Centrale de Proximité Paris-Centre". This work has been supported by the Region Ile-de-France in the framework of DIM Nano-K.

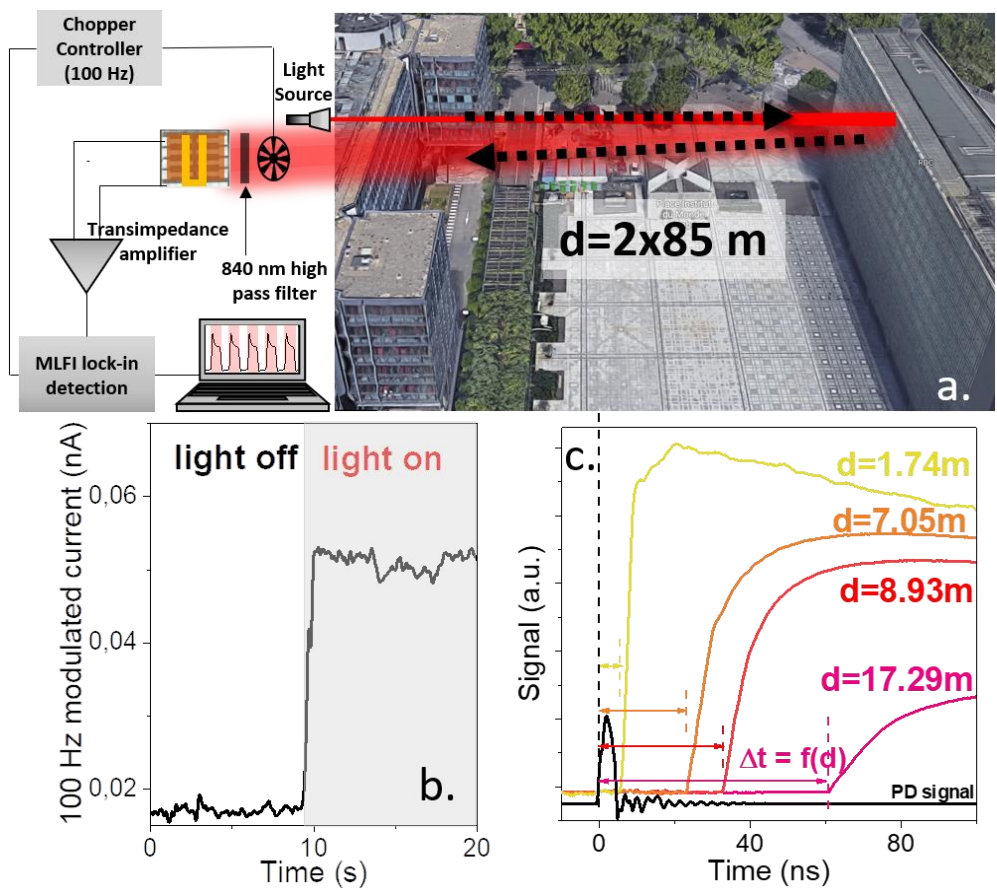

Figure 7. (a) Scheme of the long-range active imaging. A light source (10 W/940 nm lamp) enlights a building at 85 m distance. The device detects retrodiffused light from the building. (b) Photocurrent modulation measured at 100 Hz as a function of time in a long range detection configuration (c) Device response at 0 V as a function of the total distance from the source. The bottom black curve corresponds to the silicon photodiode (PD) response used for triggering. Adapted with permission from ref [83]. Copyright (2019) American Chemical Society.

### REFERENCES

1. Rogalski, A. (2017) Next decade in infrared detectors. *Electro-Optical and Infrared Systems: Technology and Applications XIV*, **10433**, 104330L.
2. Koppens, F.H.L., Mueller, T., Avouris, P., Ferrari, A.C., Vitiello, M.S., and Polini, M. (2014) Photodetectors based on graphene, other two-dimensional materials and hybrid systems. *Nature Nanotechnology*, **9** (10), 780–793.
3. Rogalski, A., Kopytko, M., and Martyniuk, P. (2019) Two-dimensional infrared and terahertz detectors: Outlook and status. *Applied Physics Reviews*, **6** (2), 021316.
4. Amani, M., Regan, E., Bullock, J., Ahn, G.H., and Javey, A. (2017) Mid-Wave Infrared Photoconductors Based on Black Phosphorus-Arsenic Alloys. *ACS Nano*, **11** (11), 11724–11731.
5. Guo, Q., Pospischil, A., Bhuiyan, M., Jiang, H., Tian, H., Farmer, D., Deng, B., Li, C., Han, S.-J., Wang, H., Xia, Q., Ma, T.-P., Mueller, T., and Xia, F. (2016) Black Phosphorus Mid-Infrared Photodetectors with High Gain. *Nano Lett.*, **16** (7), 4648–4655.
6. Keuleyan, S., Lhuillier, E., Brajuskovic, V., and Guyot-Sionnest, P. (2011) Mid-infrared HgTe colloidal quantum dot photodetectors. *Nature Photonics*, **5** (8), 489–493.
7. Zhu, B., Chen, M., Zhu, Q., Zhou, G., Abdelazim, N.M., Zhou, W., Kershaw, S.V., Rogach, A.L., Zhao, N., and Tsang, H.K. (2019) Integrated Plasmonic Infrared Photodetector Based on Colloidal HgTe Quantum Dots. *Advanced Materials Technologies*, **4** (10), 1900354.
8. Tang, X., Ackerman, M.M., Chen, M., and Guyot-Sionnest, P. (2019) Dual-band infrared imaging using stacked colloidal quantum dot photodiodes. *Nature Photonics 2019*, **13** (4), 277–282.
9. Goubet, N., Jagtap, A., Livache, C., Martinez, B., Portalès, H., Xu, X.Z., Lobo, R.P.S.M., Dubertret, B., and Lhuillier, E. (2018) Terahertz HgTe Nanocrystals: Beyond Confinement. *Journal of the American Chemical Society*, **140** (15), 5033–5036.

10. Semonin, O.E., Luther, J.M., Choi, S., Chen, H.-Y., Gao, J., Nozik, A.J., and Beard, M.C. (2011) Peak External Photocurrent Quantum Efficiency Exceeding 100% via MEG in a Quantum Dot Solar Cell. *Science*, **334** (6062), 1530–1533.
11. Ellingson, R.J., Beard, M.C., Johnson, J.C., Yu, P., Micic, O.I., Nozik, A.J., Shabaev, A., and Efros, A.L. (2005) Highly Efficient Multiple Exciton Generation in Colloidal PbSe and PbS Quantum Dots. *Nano Lett.*, **5** (5), 865–871.
12. McDonald, S.A., Konstantatos, G., Zhang, S., Cyr, P.W., Klem, E.J.D., Levina, L., and Sargent, E.H. (2005) Solution-processed PbS quantum dot infrared photodetectors and photovoltaics. *Nature Materials*, **4** (2), 138–142.
13. Kovalenko, M.V., Kaufmann, E., Pachinger, D., Roither, J., Huber, M., Stangl, J., Hesser, G., Schäffler, F., and Heiss, W. (2006) Colloidal HgTe Nanocrystals with Widely Tunable Narrow Band Gap Energies: From Telecommunications to Molecular Vibrations. *J. Am. Chem. Soc.*, **128** (11), 3516–3517.
14. Lhuillier, E., and Guyot-Sionnest, P. (2017) Recent Progresses in Mid Infrared Nanocrystal Optoelectronics. *IEEE Journal of Selected Topics in Quantum Electronics*, **23** (5), 1–8.
15. Lu, H., Carroll, G.M., Neale, N.R., and Beard, M.C. (2019) Infrared Quantum Dots: Progress, Challenges, and Opportunities. *ACS Nano*, **13** (2), 939–953.
16. Hafiz, S.B., Scimeca, M., Sahu, A., and Ko, D.-K. (2019) Colloidal quantum dots for thermal infrared sensing and imaging. *Nano Convergence*, **6** (1), 7.
17. Livache, C., Martinez, B., Goubet, N., Ramade, J., and Lhuillier, E. (2018) Road Map for Nanocrystal Based Infrared Photodetectors. *Frontiers in Chemistry*, **6**, 575.
18. Guyot-Sionnest, P., Ackerman, M.M., and Tang, X. (2019) Colloidal quantum dots for infrared detection beyond silicon. *The Journal of Chemical Physics*, **151** (6), 060901.
19. Cryer, M.E., and Halpert, J.E. (2018) 300 nm Spectral Resolution in the Mid-Infrared with Robust, High Responsivity Flexible Colloidal Quantum Dot Devices at Room Temperature. *ACS Photonics*, **5** (8), 3009–3015.
20. Chen, M., Yu, H., Kershaw, S.V., Xu, H., Gupta, S., Hetsch, F., Rogach, A.L., and Zhao, N. (2014) Fast, Air-Stable Infrared Photodetectors based on Spray-Deposited Aqueous HgTe Quantum Dots. *Advanced Functional Materials*, **24** (1), 53–59.
21. Luther, J.M., Law, M., Beard, M.C., Song, Q., Reese, M.O., Ellingson, R.J., and Nozik, A.J. (2008) Schottky Solar Cells Based on Colloidal Nanocrystal Films. *Nano Lett.*, **8** (10), 3488–3492.
22. Carey, G.H., Abdelhady, A.L., Ning, Z., Thon, S.M., Bakr, O.M., and Sargent, E.H. (2015) Colloidal Quantum Dot Solar Cells. *Chem. Rev.*, **115** (23), 12732–12763.
23. Pietryga, J.M., Schaller, R.D., Werder, D., Stewart, M.H., Klimov, V.I., and Hollingsworth, J.A. (2004) Pushing the Band Gap Envelope: Mid-Infrared Emitting Colloidal PbSe Quantum Dots. *J. Am. Chem. Soc.*, **126** (38), 11752–11753.
24. Killilea, N., Wu, M., Sytnyk, M., Yousefi Amin, A.A., Mashkov, O., Spiecker, E., and Heiss, W. (2019) Pushing PbS/Metal-Halide-Perovskite Core/Epitaxial-Ligand-Shell Nanocrystal Photodetectors beyond 3 μm Wavelength. *Advanced Functional Materials*, **29** (14), 1807964.
25. Murray, C.B., Norris, D.J., and Bawendi, M.G. (1993) Synthesis and characterization of nearly monodisperse CdE (E = sulfur, selenium, tellurium) semiconductor nanocrystallites. *J. Am. Chem. Soc.*, **115** (19), 8706–8715.
26. Green, M., and Mirzai, H. (2018) Synthetic routes to mercury chalcogenide quantum dots. *J. Mater. Chem. C*, **6** (19), 5097–5112.
27. Kershaw, S.V., Susha, A.S., and Rogach, A.L. (2013) Narrow bandgap colloidal metal chalcogenide quantum dots: Synthetic methods, heterostructures, assemblies, electronic and infrared optical properties. *Chemical Society Reviews*, **42** (7), 3033–3087.
28. Keuleyan, S., Lhuillier, E., and Guyot-Sionnest, P. (2011) Synthesis of colloidal HgTe quantum dots for narrow mid-IR emission and detection. *Journal of the American Chemical Society*, **133** (41), 16422–16424.
29. Izquierdo, E., Robin, A., Keuleyan, S., Lequeux, N., Lhuillier, E., and Ithurria, S. (2016) Strongly Confined HgTe 2D Nanoplatelets as Narrow Near-Infrared Emitters. *Journal of the American Chemical Society*, **138** (33), 10496–10501.
30. Livache, C., Izquierdo, E., Martinez, B., Dufour, M., Pierucci, D., Keuleyan, S., Cruguel, H., Becerra, L., Fave, J.L., Aubin, H., Ouerghi, A., Lacaze, E., Silly, M.G., Dubertret, B., Ithurria, S., and Lhuillier, E. (2017) Charge Dynamics and Optoelectronic Properties in HgTe Colloidal Quantum Wells. *Nano Letters*, **17** (7), 4067–4074.
31. Nasilowski, M., Mahler, B., Lhuillier, E., Ithurria, S., and Dubertret, B. (2016) Two-Dimensional Colloidal Nanocrystals. *Chemical Reviews*, **116** (18), 10934–10982.
32. Lhuillier, E., Pedetti, S., Ithurria, S., Nadal, B., Heuclin, H., and Dubertret, B. (2015) Two-Dimensional colloidal metal chalcogenides semiconductors: Synthesis, spectroscopy, and applications. *Accounts of Chemical Research*, **48** (1), 22–30.

33. Gréboval, C., Izquierdo, E., Livache, C., Martinez, B., Dufour, M., Goubet, N., Moghaddam, N., Qu, J., Chu, A., Ramade, J., Aubin, H., Cruguel, H., Silly, M., Lhuillier, E., and Ithurria, S. (2019) Impact of dimensionality and confinement on the electronic properties of mercury chalcogenide nanocrystals. *Nanoscale*, **11** (9), 4061–4066.
34. Izquierdo, E., Dufour, M., Chu, A., Livache, C., Martinez, B., Amelot, D., Patriarche, G., Lequeux, N., Lhuillier, E., and Ithurria, S. (2018) Coupled HgSe colloidal quantum wells through a tunable barrier: A strategy to uncouple optical and transport band gap. *Chemistry of Materials*, **30** (12), 4065–4072.
35. Rinnerbauer, V., Hingerl, K., Kovalenko, M., and Heiss, W. (2006) Effect of quantum confinement on higher transitions in HgTe nanocrystals. *Appl. Phys. Lett.*, **89** (19), 193114.
36. Lhuillier, E., Scarafagio, M., Hease, P., Nadal, B., Aubin, H., Xu, X.Z., Lequeux, N., Patriarche, G., Ithurria, S., and Dubertret, B. (2016) Infrared Photodetection Based on Colloidal Quantum-Dot Films with High Mobility and Optical Absorption up to THz. *Nano Letters*, **16** (2), 1282–1286.
37. Goubet, N., Livache, C., Martinez, B., Xu, X.Z., Ithurria, S., Royer, S., Cruguel, H., Patriarche, G., Ouerghi, A., Silly, M., Dubertret, B., and Lhuillier, E. (2018) Wave-Function Engineering in HgSe/HgTe Colloidal Heterostructures to Enhance Mid-infrared Photoconductive Properties. *Nano Letters*, **18** (7), 4590–4597.
38. Shen, G., and Guyot-Sionnest, P. (2019) HgTe/CdTe and HgSe/CdX (X = S, Se, and Te) Core/Shell Mid-Infrared Quantum Dots. *Chem. Mater.*, **31** (1), 286–293.
39. Sagar, L.K., Walravens, W., Maes, J., Geiregat, P., and Hens, Z. (2017) HgSe/CdE (E = S, Se) Core/Shell Nanocrystals by Colloidal Atomic Layer Deposition. *J. Phys. Chem. C*, **121** (25), 13816–13822.
40. Nag, A., Sapra, S., Nagamani, C., Sharma, A., Pradhan, N., Bhat, S.V., and Sarma, D.D. (2007) A Study of $Mn^{2+}$ Doping in CdS Nanocrystals. *Chem. Mater.*, **19** (13), 3252–3259.
41. Knowles, K.E., Hartstein, K.H., Kilburn, T.B., Marchioro, A., Nelson, H.D., Whitham, P.J., and Gamelin, D.R. (2016) Luminescent Colloidal Semiconductor Nanocrystals Containing Copper: Synthesis, Photophysics, and Applications. *Chem. Rev.*, **116** (18), 10820–10851.
42. Sahu, A., Kang, M.S., Kompch, A., Notthoff, C., Wills, A.W., Deng, D., Winterer, M., Frisbie, C.D., and Norris, D.J. (2012) Electronic Impurity Doping in CdSe Nanocrystals. *Nano Lett.*, **12** (5), 2587–2594.
43. Yang, H., Wong, E., Zhao, T., Lee, J.D., Xin, H.L., Chi, M., Fleury, B., Tang, H.-Y., Gaulding, E.A., Kagan, C.R., and Murray, C.B. (2018) Charge Transport Modulation in PbSe Nanocrystal Solids by $Au_xAg_{1-x}$ Nanoparticle Doping. *ACS Nano*, **12** (9), 9091–9100.
44. Luther, J.M., Jain, P.K., Ewers, T., and Alivisatos, A.P. (2011) Localized surface plasmon resonances arising from free carriers in doped quantum dots. *Nature Materials*, **10** (5), 361–366.
45. Deng, Z., Jeong, K.S., and Guyot-Sionnest, P. (2014) Colloidal Quantum Dots Intraband Photodetectors. *ACS Nano*, **8** (11), 11707–11714.
46. Mahler, B., Guillemot, L., Bossard-Giannesini, L., Ithurria, S., Pierucci, D., Ouerghi, A., Patriarche, G., Benbalagh, R., Lacaze, E., Rochet, F., and Lhuillier, E. (2016) Metallic Functionalization of CdSe 2D Nanoplatelets and Its Impact on Electronic Transport. *J. Phys. Chem. C*, **120** (23), 12351–12361.
47. Lee, J.-S., Shevchenko, E.V., and Talapin, D.V. (2008) Au−PbS Core−Shell Nanocrystals: Plasmonic Absorption Enhancement and Electrical Doping via Intra-particle Charge Transfer. *J. Am. Chem. Soc.*, **130** (30), 9673–9675.
48. Koh, W., Koposov, A.Y., Stewart, J.T., Pal, B.N., Robel, I., Pietryga, J.M., and Klimov, V.I. (2013) Heavily doped n- type PbSe and PbS nanocrystals using ground-state charge transfer from cobaltocene. *Scientific Reports*, **3** (1), 1–8.
49. Martinez, B., Livache, C., Meriggio, E., Xu, X.Z., Cruguel, H., Lacaze, E., Proust, A., Ithurria, S., Silly, M.G., Cabailh, G., Volatron, F., and Lhuillier, E. (2018) Polyoxometalate as Control Agent for the Doping in HgSe Self-Doped Nanocrystals. *J. Phys. Chem. C*, **122** (46), 26680–26685.
50. Miller, E.M., Kroupa, D.M., Zhang, J., Schulz, P., Marshall, A.R., Kahn, A., Lany, S., Luther, J.M., Beard, M.C., Perkins, C.L., and van de Lagemaat, J. (2016) Revisiting the Valence and Conduction Band Size Dependence of PbS Quantum Dot Thin Films. *ACS Nano*, **10** (3), 3302–3311.
51. Kroupa, D.M., Vörös, M., Brawand, N.P., McNichols, B.W., Miller, E.M., Gu, J., Nozik, A.J., Sellinger, A., Galli, G., and Beard, M.C. (2017) Tuning colloidal quantum dot band edge positions through solution-phase surface chemistry modification. *Nature Communications*, **8** (1), 15257.
52. Robin, A., Livache, C., Ithurria, S., Lacaze, E., Dubertret, B., and Lhuillier, E. (2016) Surface Control of Doping in Self-Doped Nanocrystals. *ACS Applied Materials and Interfaces*, **8** (40), 27122–27128.
53. Brown, P.R., Kim, D., Lunt, R.R., Zhao, N., Bawendi, M.G., Grossman, J.C., and Bulović, V.

(2014) Energy Level Modification in Lead Sulfide Quantum Dot Thin Films through Ligand Exchange. *ACS Nano*, **8** (6), 5863–5872.
54. Martinez, B., Livache, C., Notemgnou Mouafo, L.D., Goubet, N., Keuleyan, S., Cruguel, H., Ithurria, S., Aubin, H., Ouerghi, A., Doudin, B., Lacaze, E., Dubertret, B., Silly, M.G., Lobo, R.P.S.M., Dayen, J.F., and Lhuillier, E. (2017) HgSe Self-Doped Nanocrystals as a Platform to Investigate the Effects of Vanishing Confinement. *ACS Applied Materials and Interfaces*, **9** (41), 36173–36180.
55. Jeong, K.S., Deng, Z., Keuleyan, S., Liu, H., and Guyot-Sionnest, P. (2014) Air-Stable n-Doped Colloidal HgS Quantum Dots. *J. Phys. Chem. Lett.*, **5** (7), 1139–1143.
56. Livache, C., Martinez, B., Greboval, C., and Lhuillier, E. (2019) A Colloidal Quantum Dot Infrared Photodetector and its use for Intraband Detection. *Nature Communications*, **10** (1), 2125.
57. Shen, G., and Guyot-Sionnest, P. (2016) HgS and HgS/CdS Colloidal Quantum Dots with Infrared Intraband Transitions and Emergence of a Surface Plasmon. *J. Phys. Chem. C*, **120** (21), 11744–11753.
58. Man, P., and Pan, D.S. (1991) Infrared absorption in HgTe. *Phys. Rev. B*, **44** (16), 8745–8758.
59. Keuleyan, S.E., Guyot-Sionnest, P., Delerue, C., and Allan, G. (2014) Mercury Telluride Colloidal Quantum Dots: Electronic Structure, Size-Dependent Spectra, and Photocurrent Detection up to 12 µm. *ACS Nano*, **8** (8), 8676–8682.
60. Lhuillier, E., Keuleyan, S., and Guyot-Sionnest, P. (2012) Optical properties of HgTe colloidal quantum dots. *Nanotechnology*, **23** (17), 175705.
61. Jagtap, A.M., Martinez, B., Goubet, N., Chu, A., Livache, C., Greboval, C., Ramade, J., Amelot, D., Trousset, P., Triboulin, A., Ithurria, S., Silly, M.G., Dubertret, B., Lhuillier, E., Gréboval, C., Ramade, J., Amelot, D., Trousset, P., Triboulin, A., Ithurria, S., Silly, M.G., Dubertret, B., and Lhuillier, E. (2018) Design of Unipolar Barrier for Nanocrystal Based Short Wave Infrared Photodiode. *ACS Photonics*, **5** (11), 4569–4576.
62. Chu, A., Martinez, B., Ferré, S., Noguier, V., Gréboval, C., Livache, C., Qu, J., Prado, Y., Casaretto, N., Goubet, N., Cruguel, H., Dudy, L., Silly, M.G., Vincent, G., and Lhuillier, E. (2019) HgTe Nanocrystals for SWIR Detection and Their Integration up to the Focal Plane Array. *ACS Appl. Mater. Interfaces*, **11** (36), 33116–33123.
63. Chen, M., and Guyot-Sionnest, P. (2017) Reversible Electrochemistry of Mercury Chalcogenide Colloidal Quantum Dot Films. *ACS Nano*, **11** (4), 4165–4173.
64. Chuang, C.-H.M., Brown, P.R., Bulović, V., and Bawendi, M.G. (2014) Improved performance and stability in quantum dot solar cells through band alignment engineering. *Nature Materials*, **13** (8), 796–801.
65. Lhuillier, E., Keuleyan, S., Zolotavin, P., and Guyot-Sionnest, P. (2013) Mid-infrared HgTe/$As_2S_3$ field effect transistors and photodetectors. *Advanced Materials*, **25** (1), 137–141.
66. Kovalenko, M.V., Scheele, M., and Talapin, D.V. (2009) Colloidal Nanocrystals with Molecular Metal Chalcogenide Surface Ligands. *Science*, **324** (5933), 1417–1420.
67. Martinez, B., Ramade, J., Livache, C., Goubet, N., Chu, A., Gréboval, C., Qu, J., Watkins, W.L., Becerra, L., Dandeu, E., Fave, J.L., Méthivier, C., Lacaze, E., and Lhuillier, E. (2019) HgTe Nanocrystal Inks for Extended Short-Wave Infrared Detection. *Advanced Optical Materials*, 1900348.
68. Chen, M., Lan, X., Tang, X., Wang, Y., Hudson, M.H., Talapin, D.V., and Guyot-Sionnest, P. (2019) High Carrier Mobility in HgTe Quantum Dot Solids Improves Mid-IR Photodetectors. *ACS Photonics*, **11**, 48.
69. Livache, C., Goubet, N., Martinez, B., Jagtap, A., Qu, J., Ithurria, S., Silly, M.G., Dubertret, B., and Lhuillier, E. (2018) Band Edge Dynamics and Multiexciton Generation in Narrow Band Gap HgTe Nanocrystals. *ACS Applied Materials and Interfaces*, **10** (14), 11880–11887.
70. Yifat, Y., Ackerman, M., and Guyot-Sionnest, P. (2017) Mid-IR colloidal quantum dot detectors enhanced by optical nano-antennas. *Appl. Phys. Lett.*, **110** (4), 041106.
71. Le-Van, Q., Le Roux, X., Aassime, A., and Degiron, A. (2016) Electrically driven optical metamaterials. *Nature Communications*, **7** (1), 12017.
72. Baek, S.-W., Ouellette, O., Jo, J.W., Choi, J., Seo, K.-W., Kim, J., Sun, B., Lee, S.-H., Choi, M.-J., Nam, D.-H., Quan, L.N., Kang, J., Hoogland, S., García de Arquer, F.P., Lee, J.-Y., and Sargent, Edward.H. (2018) Infrared Cavity-Enhanced Colloidal Quantum Dot Photovoltaics Employing Asymmetric Multilayer Electrodes. *ACS Energy Lett.*, **3** (12), 2908–2913.
73. Tang, X., fu Wu, G., and Lai, K.W.C. (2017) Plasmon resonance enhanced colloidal HgSe quantum dot filterless narrowband photodetectors for mid-wave infrared. *Journal of Materials Chemistry C*, **5** (2), 362–369.
74. Tang, X., Ackerman, M.M., and Guyot-Sionnest, P. (2018) Thermal Imaging with Plasmon Resonance Enhanced HgTe Colloidal Quantum Dot Photovoltaic Devices. *ACS Nano*, **12** (7), 7362–7370.

75. Prins, F., Kim, D.K., Cui, J., De Leo, E., Spiegel, L.L., McPeak, K.M., and Norris, D.J. (2017) Direct Patterning of Colloidal Quantum-Dot Thin Films for Enhanced and Spectrally Selective Out-Coupling of Emission. *Nano Lett.*, **17** (3), 1319–1325.
76. De Leo, E., Cocina, A., Tiwari, P., Poulikakos, L.V., Marqués-Gallego, P., le Feber, B., Norris, D.J., and Prins, F. (2017) Polarization Multiplexing of Fluorescent Emission Using Multiresonant Plasmonic Antennas. *ACS Nano*, **11** (12), 12167–12173.
77. Baek, S.-W., Molet, P., Choi, M.-J., Biondi, M., Ouellette, O., Fan, J., Hoogland, S., Arquer, F.P.G. de, Mihi, A., and Sargent, E.H. (2019) Nanostructured Back Reflectors for Efficient Colloidal Quantum-Dot Infrared Optoelectronics. *Advanced Materials*, **31** (33), 1901745.
78. Chu, A., Gréboval, C., Goubet, N., Martinez, B., Livache, C., Qu, J., Rastogi, P., Bresciani, F.A., Prado, Y., Suffit, S., Ithurria, S., Vincent, G., and Lhuillier, E. (2019) Near Unity Absorption in Nanocrystal Based Short Wave Infrared Photodetectors Using Guided Mode Resonators. *ACS Photonics*, **6** (10), 2553–2561.
79. Rauch, T., Böberl, M., Tedde, S.F., Fürst, J., Kovalenko, M.V., Hesser, G., Lemmer, U., Heiss, W., and Hayden, O. (2009) Near-infrared imaging with quantum-dot-sensitized organic photodiodes. *Nature Photonics*, **3** (6), 332–336.
80. Ciani, A.J., Pimpinella, R.E., Grein, C.H., and Guyot-Sionnest, P. (2016) Colloidal quantum dots for low-cost MWIR imaging. *Infrared Technology and Applications XLII*, **9819**, 981919.
81. Buurma, C., Pimpinella, R.E., Ciani, A.J., Feldman, J.S., Grein, C.H., and Guyot-Sionnest, P. (2016) MWIR imaging with low cost colloidal quantum dot films. *Optical Sensing, Imaging, and Photon Counting: Nanostructured Devices and Applications 2016*, **9933**, 993303.
82. Ramade, J., Qu, J., Chu, A., Greboval, C., Livache, C., Goubet, N., Martinez, B., VINCENT, G., and Lhuillier, E. (2020) Potential of Colloidal Quantum Dot based Solar Cell for Near-Infrared Active Detection. *ACS Photonics* 7(1), 272-278.